\documentclass[pre,twoside,twocolumn,superscriptaddress,floatfix]{revtex4-1}

\usepackage{graphicx,amssymb,amsmath,epsfig,color,enumerate,bm}
\usepackage{etoolbox} % for \appto

\makeatletter
\appto{\appendix}{%
  \@ifstar{\def\theequation@prefix{A.}}%
          {}%
}
\makeatother
\epsfclipon

\makeatletter
\appto{\appendix}{%
  \@ifstar{\def\theequation@prefix{A.}}%
          {}%
}
\makeatother

\begin{document}
 \title{Non-Gaussian diffusion in static disordered media}
 \author{Liang Luo}
 \author{Ming Yi}
  \email{yiming@mail.hzau.edu.cn}
\affiliation{Department of Physics, Huazhong Agricultural University, Wuhan 430070, China}
\affiliation{Institute of Applied Physics, Huazhong Agricultural University, Wuhan 430070, China}

\begin{abstract}
Non-Gaussian diffusion is commonly considered as a result of fluctuating diffusivity, which is correlated in time or in space or both. In this work, we investigate the non-Gaussian diffusion in static disordered media via a quenched trap model, where the diffusivity is spatially correlated. Several unique effects due to quenched disorder are reported. We analytically estimate the diffusion coefficient $D_{\text{dis}}$ and its fluctuation over samples of finite size. We show a mechanism of population splitting in the non-Gaussian diffusion. It results in a sharp peak in the distribution of displacement $P(x,t)$ around $x=0$, that has frequently been observed in experiments. We examine the fidelity of the coarse-grained diffusion map, which is reconstructed from particle trajectories. Finally, we propose a procedure to estimate the correlation length in static disordered environments, where the information stored in the sample-to-sample fluctuation has been utilized. 
\end{abstract}

%\pacs{87.16.dj, 02.50.-r, 05.40.-a, 87.10.Mn}

\maketitle

\section{Introduction}
Modern imaging experiments provide a huge amount of dynamic details of diffusion in crowded intracellular environments\cite{barkai12}, which has greatly deepen our understanding of the underlying stochastic motion\cite{bressloff13}. A novel class of anomalous diffusion, the non-Gaussian diffusion, has been frequently reported by experiments in a wide range of disordered systems, including soft matter systems\cite{wang09,wang12,kou17}, cytoplasm\cite{stylianidou14,li15,munder16}, cell membrane\cite{he16,jeon16}, and even in the heat transport problem\cite{zhao16}. 

In the non-Gaussian diffusion, the distribution of displacement is not Gaussian, 
while the mean squared displacement can be either linear\cite{wang09,wang12,he16,jeon16} or sub-linear\cite{kou17,stylianidou14,li15,munder16,ghosh16} to time. A simple interpretation suggests the dynamic heterogeneity plays a key role in this phenomenon\cite{wang09,wang12}. In the case that each particle diffuses with random instantaneous diffusivity $D^{(t)}$, the statistics over all the possible diffusivity introduces a convolution to the distribution of displacement by 
\begin{equation}
P(x,t)=\int_0^{\infty}dD^{(t)}G(x,t\vert D^{(t)})P(D^{(t)}). 
\end{equation}
In general, $P(x,t)$ is non-Gaussian even when $G(x,t\vert D^{(t)})$ is Gaussian. 
A theory of fluctuating diffusivity was formulated\cite{metzler13,slater14,uneyama15,akimoto16pre,miyaguchi16,bodrova16, cherstvy16, aurell17}, where the diffusivity of each particle follows an independent stochastic process. The theory has been supported by direct observation of fluctuating diffusivity in experiments\cite{he16,manzo15} and simulation\cite{jeon16}. Most recently, a comprehensive theoretical framework for the random walk of fluctuating diffusivity was constructed, based on the idea of subordination\cite{metzler17}.

It is a general concern that the dynamic heterogeneity may be introduced by the quasi-static disordered environment \cite{munder16,jeon16,slater14,ghosh15,ye12}, where the diffusivity is correlated in space instead of in time. Our understanding on non-Gaussian diffusion in a static environment, however, is still very limited. We note such anomalous diffusive processes can be described by the random walk on the lattices of quenched traps\cite{haus87}, which has been intensively studied in the context of sub-diffusion\cite{bouchaud90,monthus03,bertin03,barkai11,luo14}. The quenched trap model provides insights into the fluctuation among disordered static samples\cite{luo15,akimoto16}, which is essential in biology, known as ``every cell is different''. 

In the current work, we study a quenched trap model for non-Gaussian diffusion\cite{luo16}, where the landscape is locally 
correlated and the local diffusivity $D^{(l)}$ follows the exponential distribution $P(D^{(l)}=D)=D_0^{-1}\exp(-D/D_0)$. Three effects due to the quenched disorder are reported in this paper. 
(1) A sharp peak due to population splitting arises in the distribution of displacement $P(x,t)$ around $x=0$. 
(2) In the case that $0<P(D\simeq0)<\infty$, the measured diffusion coefficient depends on the sample size. 
(3) The coarse-grained diffusion map reconstructed from the trajectories is faithful to the genuine landscape only when the spatial resolution is high enough to distinguish the fine structures of the local domains. Inspired by the coarse-graining processes, we propose an approach to estimate the correlation length in samples. 

The paper is organized as follows. In Sec.\ref{sec_model} we introduce the quenched trap model. Section \ref{sec_ue} reports unique effects due to static disordered environments. Section \ref{sec_da} provides insights to the trajectory-based data analysis. We discuss generalizations of the model and its connection to other works in Sec.\ref{sec_ds}. Finally, we give a brief summary in Sec. \ref{sec_sm}. 

\section{Quenched trap model}
\label{sec_model}

Let us begin with the simplest form of the quenched trap model, 
i.e., a particle hopping on a $d$-dimensional simple cubic lattice 
with a set of site-dependent transition rates $W_{i\rightarrow j}=n_c^{-1}\tau_i^{-1}$ from site $i$ to one of its nearest 
neighbours $j$. Here $n_c=2d$ is the lattice coordination number. The hopping rate $k_i=\tau_i^{-1}$ out of site $i$
can be associated with a site energy $V_i (<0)$ through the Arrhenius law $k_i=\omega_0\exp(V_i/T)$, where 
$\omega_0$ is the attempt rate and $T$ the ambient temperature. The local diffusivity can be defined for each site by
\begin{equation}
\label{eq_dv}
D_i^{(l)}\equiv\frac{a^2}{2d\tau_i}=\frac{\omega_0 a^2}{2d}e^{V_i/T},
\end{equation}
where $a$ is the lattice constant and the superscript $(l)$ indicates 'local'. We set $\omega_0=1, a=1$ for convenience and consider the two-dimensional case $d=2$. 

\begin{figure}
\centering
\includegraphics[width=8.6cm]{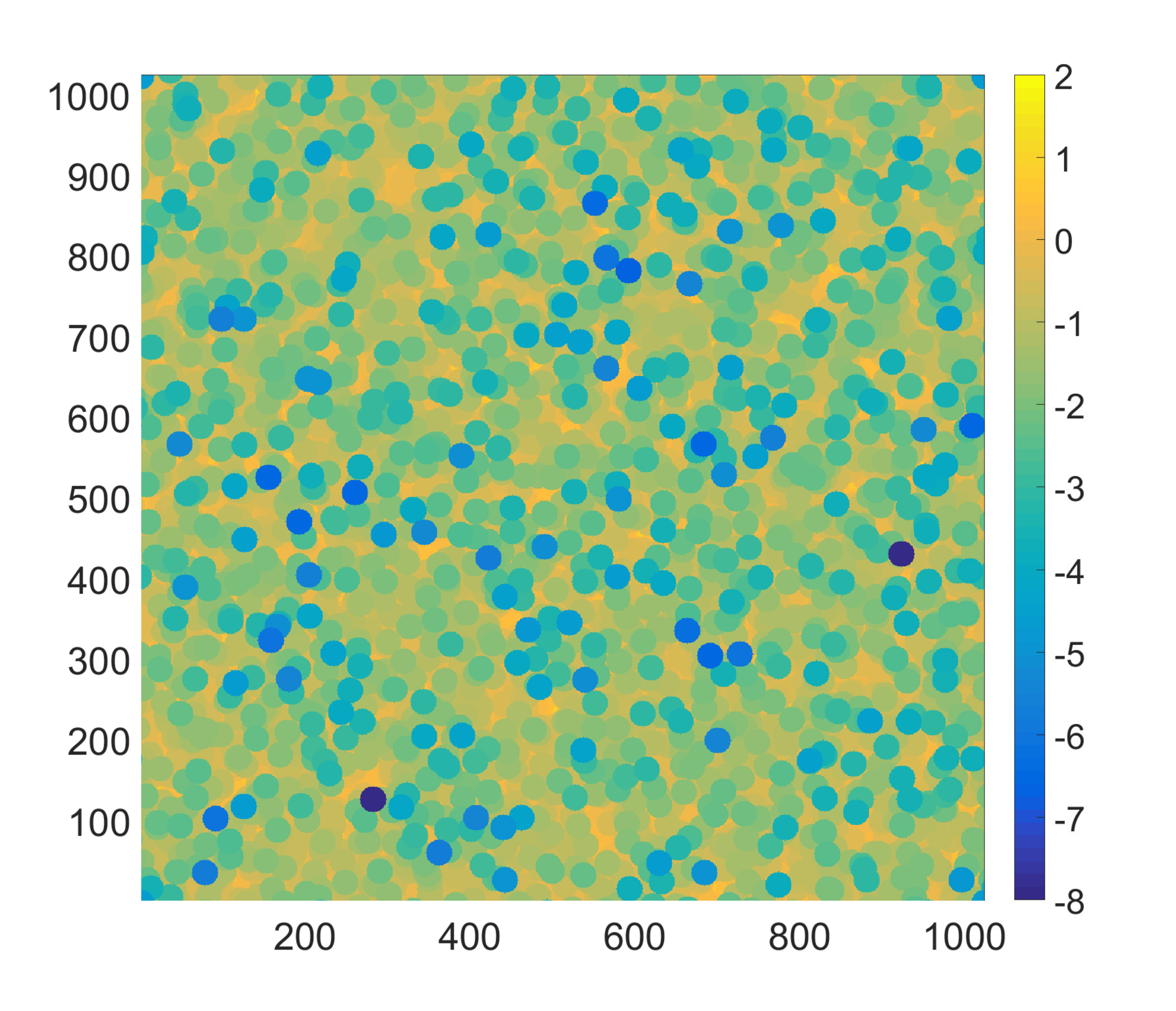}
\caption{\label{fig_dmap} The color map of a typical sample of  correlated random energies with $L=1024$ and $r_c=16$. The color is assigned to each site according to $V_i=\ln D^{(l)}_i$, where $D_0=0.25$ is chosen. }
\end{figure}

In this work, we focus on a special random energy landscape $\{V_i\}$, 
of  which the diffusivity is locally correlated and follows the exponential distribution
\begin{equation}
\label{eq_pd}
P(D^{(l)}_i=D)=D_0^{-1}\exp\left(-D/D_0\right),\qquad(D>0).
\end{equation}
The spatial correlation is usually introduced by independent modes in Fourier space. 
This approach is not convenient here, however, since it always leads to Gaussian distributed random variables. 
We note that corresponding to $P(D)$ given by Eq.(\ref{eq_pd}), $\{V_i=\ln D^{(l)}_i\}$ follows the Gumbel distribution
\begin{equation}
\label{eq_gumbel}
P(V_i=V)=\exp\left[V-\ln D_0-\exp\left(V-\ln D_0\right)\right],
\end{equation}
which is the limiting distribution of extreme statistics. 
A two-step approach based on extreme statistics is hence introduced for the spatially correlated $\{V_i\}$. 
First generate an uncorrelated landscape $\{U_i\}$, following the exponential distribution
$P(U_i=U)=U_0^{-1}\exp\left(U/U_0\right)$. Then assign to $V_i$ the minimum energy in the $r_c-$neighbourhood of $i$, i.e. $V_i=\min \left\{U_j\vert r_{ij}<r_c\right\}$. Noting that $\{U_i\}$ is uncorrelated, one can see $P(V_i)$ converges to the Gumbel distribution for large $r_c$. $D^{(l)}_i=e^{V_i/T}/2d$ with $T=1$ is then exponentially distributed, which is confirmed by numerical sampling for $r_c=16$.

The extreme landscape $\{V_i\}$ is composed of disks of local extreme values, overlapping with each other, as shown in Fig.\ref{fig_dmap}. The spatial correlation of $\{D^{(l)}_i\}$ is hence introduced up to the basin size $\xi\approx2r_c$. The deepest basins are of full shape as the whole disks,  shown as the blue ones in the color map. The more shallow basins are frequently overlapped by the neighbour ones since larger density of states. They hence appear as smaller but denser pieces of disks, which constitute the rather continuous part of the landscape, shown as the yellow region in the color map. 

As a generalization, one can generate a class of correlated landscapes of $\{D^{(l)}_i\}$ from given $\{V_i\}$ via  Eq.(\ref{eq_dv}) with various $T\neq 1$.  Noting $P(D^{(l)}_i=D)dD=P(V_i=V)dV$, it can be shown that $P(D^{(l)}_i=D)= D_1^{-1}TD^{T-1}e^{-D^T/D_1}$, where $D_1=D_0/(2d)^T$. Let us focus on the $T=1$ case in this work. 

\section{Unique effects due to static disordered environments}
\label{sec_ue}

\subsection{Sample-dependent diffusion coefficient}
\label{sec_dc}

For normal Brownian motion, the mean square displacement (MSD) $\left<x^2\right>=\left<\vert x(t)-x(0)\vert^2\right>$ is expected linear to time $t$ in long time limit. The diffusion coefficient is defined by $D_{\text{dis}}\equiv\lim_{t\rightarrow\infty}\left<x^2\right>/4t$, 
where the average $\left<\cdot\right>$ is performed over the trajectories. 
%We would like to notice here we introduce three measures in this work, the diffusion coefficient $D_{\text{dis}}$ estimated by the above means, the instantaneous diffusivity $D^{(t)}$ measured from short segments of trajectories, and the local diffusivity $D^{(l)}$ fixed in the static samples. 
Following Kehr and Haus \cite{haus87}, the diffusion coefficient of trap model depends on the local diffusivity
\begin{equation}
\label{eq_dtau2}
D_{\text{dis}}=\left[\frac{1}{N}\sum_{i=1}^N (D_i^{(l)})^{-1}\right]^{-1}. 
\end{equation}
It is confirmed by our simulation (not shown here). 
Its relation to the instantaneous diffusivity $D^{(t)}$ is shown later by Eq.(\ref{eq_pdt0}-\ref{eq_pdt2}).
We would like to call readers' attention that $D_{\text{dis}}$ is in general different to $D_0=\lim_{N\rightarrow\infty}\frac{1}{N}\sum_{i=1}^N D_i^{(l)}$. 
One can find more details in the recent paper by Akimoto {\it et al.} \cite{akimoto16}. 
Noting $\tau_i=a^2/(2d D_i^{(l)})$, one can see the connection between the diffusion coefficient and the mean sojourn time of
the given sample
\begin{equation}
\label{eq_taubar}
\bar{\tau}_N\equiv\frac{1}{N}\sum_{i=1}^N \tau_i=\frac{a^2}{2dD_{\text{dis}}}. 
\end{equation}

\begin{figure}
\centering
\includegraphics[width=8.6cm]{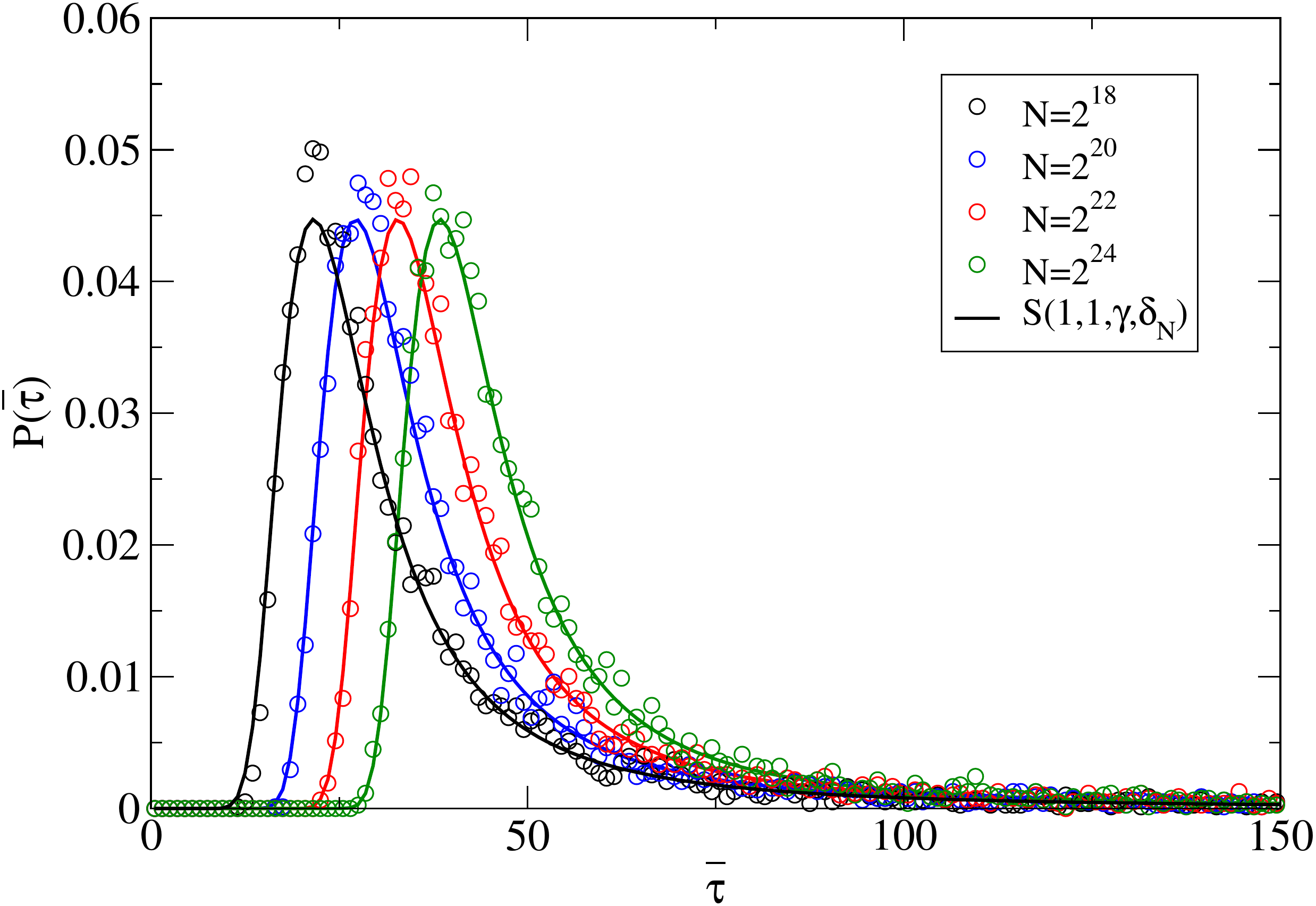}
\caption{\label{fig_ptau} The distribution of mean waiting time $\bar\tau_N$ for samples with $r_c=16$, $D_0=0.25$ and various sizes $N=L^2$. $8000$ landscapes are sampled for each size. The solid lines are the probability density function of one-sided L\`{e}vy stable distribution with $\mu=1$, $\gamma=6.35$, and $\delta_N=\delta_0+\frac{2}{\pi}\gamma\ln N$ with $\delta_0=-26.1$. 
}
\end{figure}

$D_{\text{dis}}$ and $\bar{\tau}$ depend on the specific configuration of the sample. We further consider the statistics over the ensemble of samples, first estimating the distribution of $\bar{\tau}$, then the mean value and higher moments of $D_{\text{dis}}$. Exact estimation is challenging due to the spatial correlation. Noting that the correlation among extremal basins is quite weak, we consider the coarse-grained lattice of basins instead. The typical sojourn time in basin $i$ is proportional to the inverse of  the local diffusivity, $\tau_i\simeq 1/D_i^{(l)}$. Noting that $D_i^{(l)}$ follows exponential distribution given by Eq.(\ref{eq_pd}), we have
\begin{equation}
P(\tau_i=\tau)=\tau_0\tau^{-2}\exp\left(-\tau_0/\tau\right),
\end{equation}
where $\tau_0\simeq1/D_0$. 
The distribution is with heavy tail $P(\tau)\sim\tau^{-2}$, which is mainly contributed by the frozen sites with $P(D_i^{(l)}=0)=1/D_0$. The expectation value of $\tau$ diverges. 
The generalized central limit theorem suggests the mean waiting time $\bar{\tau}_N$ follows the one-sided L\'{e}vy stable distribution with the exponent $\mu=1$ and the skewness $\beta=1$, 
\begin{equation}
\label{eq_stable}
\bar{\tau}_N=\frac{1}{N}\sum_{i=1}^N\tau_i\xrightarrow{d}z \sim S(\mu=1,\beta=1, \gamma,\delta_N;1), \text{for large $N$},
 \end{equation} 
where $\xrightarrow{d}$ means converging in the sense of the probability distribution. Here we adopt the type-1 parameterization of the stable distribution, where $\gamma$ is the scale parameter and $\delta_N=\delta_0+\frac{2}{\pi}\gamma\ln N$ is the position parameter\cite{nolan}. As a consequence, the position of the distribution shifts to infinity by $\ln N$, which is well confirmed by numerical sampling (see Figure \ref{fig_ptau}). 

\begin{figure}
\centering
\includegraphics[width=8.6cm]{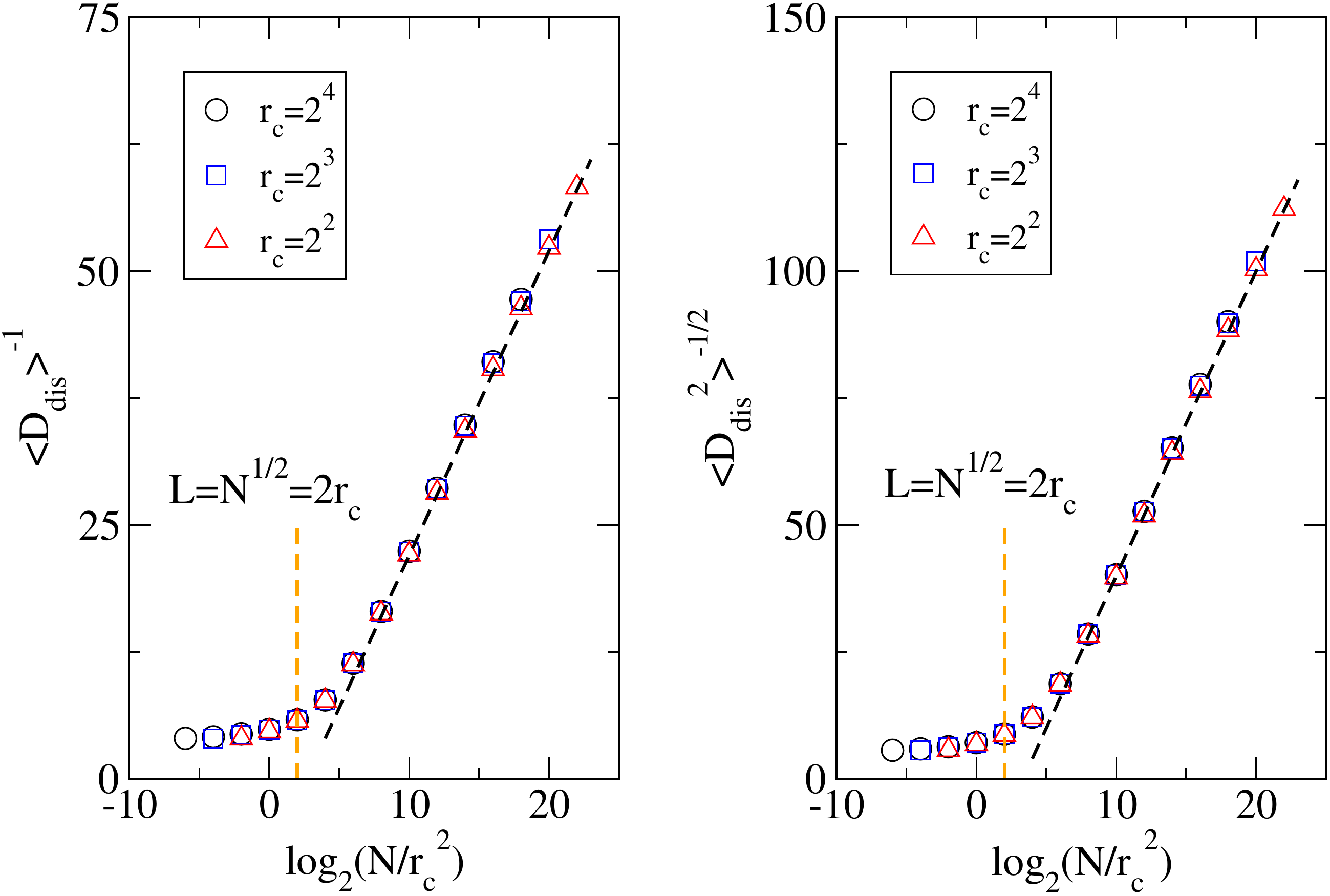}
\caption{\label{fig_dn2} $\left<D_{\text{dis}}\right>^{-1}$ (left) and $\left<D_{\text{dis}}^2\right>^{-1/2}$ (right) are plotted against $\log_2(N/r_c^2)$. The averaging is performed over $8000$ samples, for various sample sizes $N=L^2$ and various extremal basin radii $r_c$. The dash lines are added for guidance. 
}
\end{figure}

Noting that $D_{\text{dis}}=a^2/(2d\bar\tau_N)$, we see the moments of $D_{\text{dis}}$ are indeed negative moments of $\bar{\tau}_N$ 
\begin{equation}
\left<(D_{\text{dis}})^{\alpha}\right>=\left(\frac{a^2}{2d}\right)^{\alpha}\left<\bar{\tau}_N^{-\alpha}\right>.
\end{equation}
Noting the Laplace transform of the one-sided L\'{e}vy stable distribution\cite{chechkin09,uchaikin99}, it can be shown that
\begin{equation}
\label{eq_dsize}
\left<D_\text{dis}\right>^{\alpha}\sim \left(\delta_0+\frac{2}{\pi}\ln N\right)^{-\alpha}. 
\end{equation}
(See Appendix \ref{apdx_nm} for details.)
The analytical results are confirmed by numerical sampling for $\alpha=1,2$, as shown in Fig.\ref{fig_dn2}. 
One can read that the uncorrelated approximation works well for samples larger than the extremal basins, i.e. $L>\xi=2r_c$. 

\subsection{Non-Gaussian distribution of displacement with additional peak around origin}
\label{sec_ng}

For more dynamic details, we performed kinetic Monte Carlo simulation\cite{gillespie77} for the random walk on two-dimensional quenched samples with periodic boundary. The random walks are composed by hops between nearest-neighbour sites. The trapping time constants $\tau_i$ are assigned to the sites of quenched sample. The actual waiting time for a hop follows an exponential distribution with the given time constants. The summation of the waiting times gives the total walk time of a particle. The random walk is terminated when the total time reaches an upper boundary $t_{\text{max}}$, which can be understood as the limited time duration of the imaging experiment. The trajectory is further discretized by a fixed time bin $\Delta t$, to mimic the limited time resolution of the camera. $10000$ trajectories are sampled for each disordered realization. 
The initial sites of trajectories are chosen from Boltzmann distribution. Being specific, a trajectory starts from site $i$ with the probability
\begin{equation}
\label{eq_boltzmann}
P_i=\frac{e^{-V_i/T}}{Z}, 
\end{equation}
where $Z=\sum_{i=1}^{N} e^{-V_i/T}$ is the partition function over all the $N=L^2$ sites. 
In this work, the sample size is set by $L=1024$, while the radius of the extreme basin is set as $r_c=16$. 
The mean local diffusivity is set as $D_0=0.25$.  The time bin is set as $\Delta t=10$. The total observation time is set as $t_{\text{max}}=25000$, so that $n_{\text{frame}}=2500$ frames are recorded for each trajectory. 

\begin{figure}
\centering
\includegraphics[width=8.6cm]{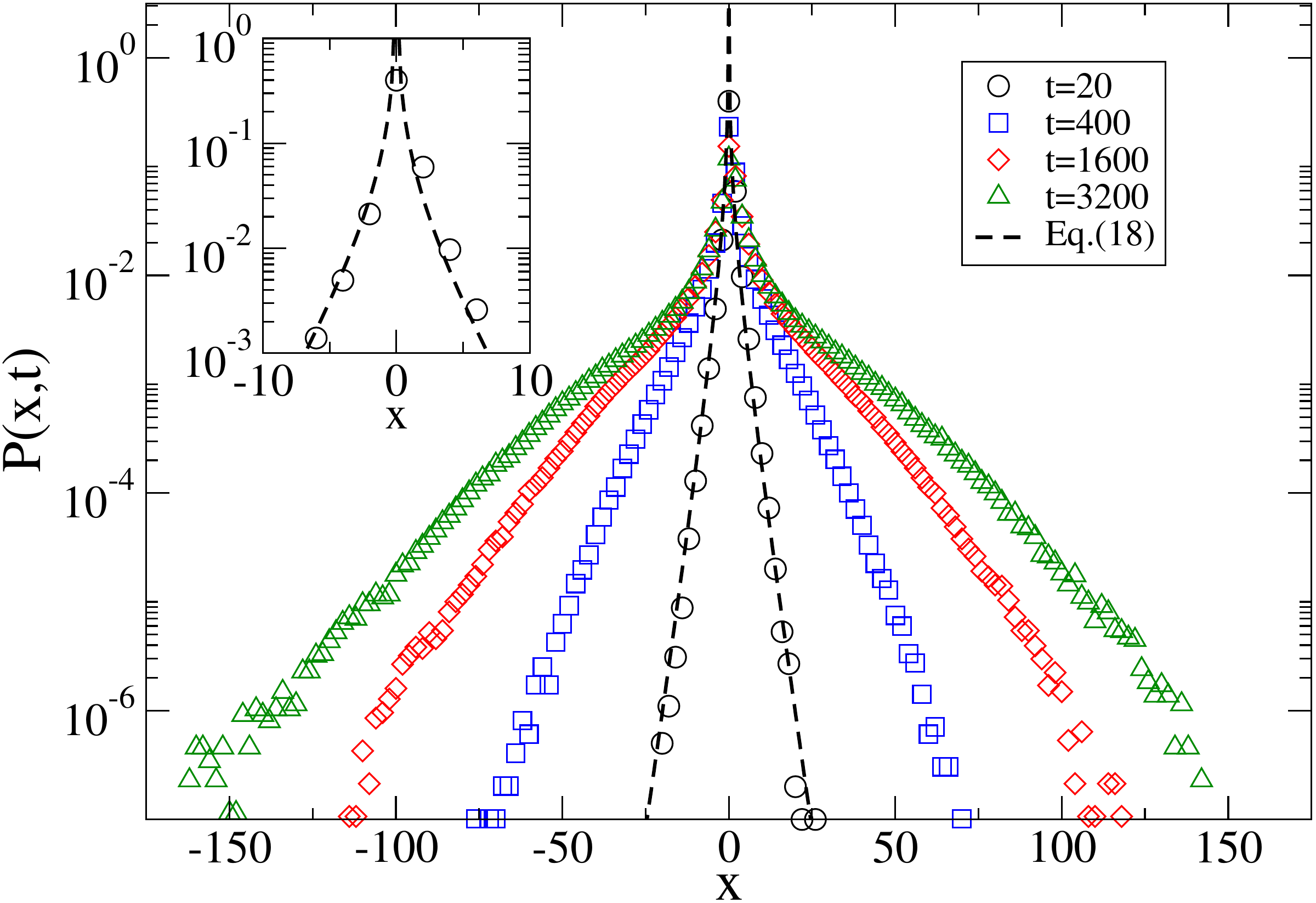}
\caption{\label{fig_ngdd}The distribution of displacement $P(x,t)$ for a typical sample and various $t$. 
The leading term of Eq.(\ref{eq_peak}) is plotted by the dash line. The inset enlarges the peak of $P(x,t=20)$ for a clearer view. 
}
\end{figure}

Figure \ref{fig_ngdd} shows the distribution of displacement $P(x,t)$ for various lag time $t$. 
As can be seen in the figure, the tail of the distribution changes continuously for increasing $t$, from the exponential tail towards Gaussian. Such behavior was also reported in previous studies\cite{slater14,jeon16,metzler17}. 
It can be well explained by noticing the heterogeneity in the ensemble of the trajectories\cite{metzler17}. 
In the early stage of each trajectory, the random walk is dominated by the local diffusivity of the initial extremal basin. The exponential distributed local diffusivity introduces the non-Gaussian tail. 
At the later stage a large fraction of particles leave the original extremal basins. Self-averaging is gradually achieved along the long trajectories of these fast moving particles, which leads to the Gaussian tail. The characteristic time scale for the convergence to the Gaussian ones has been well discussed in the annealed models of fluctuating diffusivity\cite{uneyama15,miyaguchi16,metzler17}.

A sharp peak of $P(x,t)$ appears around $x=0$ for all $t$. 
Similar peaks were also observed in various systems of non-Gaussian diffusion, 
both by experiments\cite{wang09,wang12,he16,munder16} and by molecular dynamics simulations\cite{jeon16}. 
Such peak has been studied as the phenomena of ``population splitting'' in the context of sub-diffusion\cite{barkai2003prl,barkai2003jcp,schulz13,schulz14}, where the heavy-tailed sojourn time distribution introduces localization. The immobilized particles largely influence the statistics over time ensemble. 
The peak appearing in non-Gaussian diffusion can be similarly explained in the annealed framework of fluctuating diffusivity\cite{slater14,metzler17}, but at the cost of introducing additional immobile state. 
In the quenched model, however, it is a natural consequence of the localization due to the coupling between the local diffusivity and the sojourn time in the trap. It can be simply explained as: when the particle is trapped in area where the "slow" state is preferred, to escape would be much harder, since it walks slower. The quantitative description follows. The distribution of displacement $P(x,t)$ counts all the segments $\{x(t_0)\rightarrow x(t_0+t)\}$ of the trajectories. For small $t$, we assume each segment is dominated by a fixed instantaneous diffusivity $D^{(t)}$, which is equal to the local diffusivity $D^{(l)}_i$ of the center of the segment. Noting that the segments sample the landscape with the Boltzmann weight (Eq.(\ref{eq_boltzmann})) and also Eq.(\ref{eq_dv}), we can get the probability that a segment centers at site $i$ in the condition of $D^{(l)}_i=D$ and measured $D_{\text{dis}}$ by
\begin{equation} 
\label{eq_px}
P\left(x(t)=x_i\vert D_i^{(l)}=D,D_{{\text{dis}}}\right)=\frac{D^{-1}}{\sum_{j=1}^N (D_j^{(l)})^{-1}},
\end{equation}
where the denominator $\sum_{j=1}^N (D_j^{(l)})^{-1}=ND^{-1}_{\text{dis}}$ is given by the diffusion coefficient of the specific sample. 
The instantaneous diffusivities $D^{(t)}$ of the segments hence follow the distribution with additional weight
\begin{equation}
\label{eq_pdt0}
P(D^{(t)}=D\vert D_{\text{dis}})=\sum_{i=1}^N P(x=x_i\vert D_i^{(l)}=D, D_{\text{dis}})P(D_i^{(l)}=D\vert D_{\text{dis}}).
\end{equation}
We would like to call reader's attention that all the above probabilitis are under the condition of the known $D_{\text{dis}}$ of the specific sample. Employing  Eq.(\ref{eq_pd}), Eq.(\ref{eq_dtau2}) and Eq.(\ref{eq_px}), it can be shown that $D_{\text{dis}}$ indeed gives a lower bound of $D^{(l)}$. A small-$D$ cutoff hence arises naturally, which leads to
\begin{equation}
\label{eq_pdt1}
P(D^{(t)}=D\vert D_{\text{dis}})=0,\hspace{2em}D\le D_c,
\end{equation}
where $D_c=D_{\text{dis}}/N$. 
When $D^{(l)}=D\gg D_c$, the probability $P(D^{(l)}=D)$  and $P(D_{\text{dis}})$ are independent, which leads to
\begin{equation}
\label{eq_pdt2}
P(D^{(t)}=D\vert D_{\text{dis}})=\frac{D_{\text{dis}}}{D_0}D^{-1}e^{-D/D_0}, D\gg D_c.
\end{equation}
%\begin{equation}
%\label{eq_pdt}
%  P(D^{(t)}=D\vert D_{\text{dis}})=\left\{
%   \begin{array}{rl}
%   0,\hspace{6em} D\le D_c\\
%   \frac{D_{\text{dis}}}{D_0}D^{-1}e^{-D/D_0}, D\gg D_c\\
%   \end{array}
%  \right.
%\end{equation}
One can find the technical details in Appendix \ref{apdx_dt}.
%
% &=&\frac{D_{\text{dis}}}{D_0}D^{-1}e^{-D/D_0},
%\end{eqnarray}

\begin{figure*}
\includegraphics[width=17.8cm]{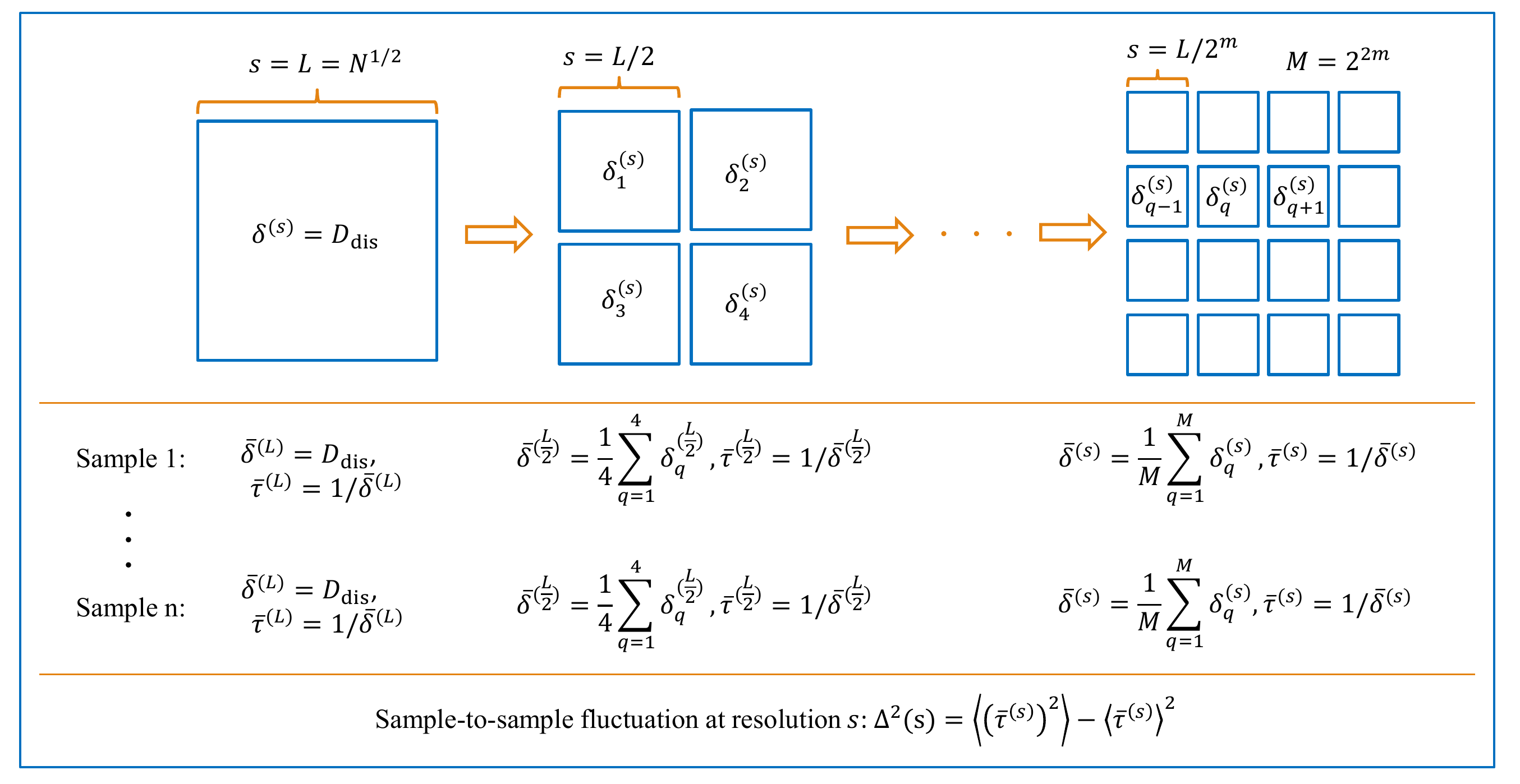}
\caption{\label{fig_scheme}A scheme for the trajectory-based data analysis, discussed in Sec.\ref{sec_da}. The sample is divided into small grains of size $s$. The local diffusivity $\delta^{(s)}_q$ is evaluated for the grain $q$ from the segments of trajectories centered therein (see Eq.(\ref{eq_ds})). The mean local diffusivity $\bar{\delta}^{(s)}$ and the typical time $\bar{\tau}^{(s)}$ can be evaluated for each sample at various resolution $s$. The sample-to-sample fluctuation of $\bar{\tau}^{(s)}$ is expected vanishing while $s$ approaches the typical correlation length $\xi$. }
\end{figure*}

Further assuming in each short segment the particle diffuses as normal Brownian motion dominated by the instantaneous diffusivity $D^{(t)}$, we have
\begin{equation}
G(x,t\vert D^{(t)})=\frac{1}{\sqrt{4\pi D^{(t)}t}}\exp\left(-\frac{x^2}{4D^{(t)}t}\right). 
\end{equation}
The distribution of displacement is obtained by counting all the segments, which gives
\begin{equation}
\label{eq_peak0}
P(x,t)=\int_{0}^{\infty}dD\;G(x,t\vert D^{(t)})P(D^{(t)}=D\vert D_{\text{dis}}).
%	&=&\frac{D_{\text{dis}}}{D_0}x^{-1}\exp(-\frac{x}{\sqrt{D_0t}}).
\end{equation}
It is fortunate that the above integral can be explicitly expressed by the Gauss error function.
For $x>x_c\equiv\sqrt{D_c t}$ and $D_c\ll D_0$, it can be shown
\begin{equation}
\label{eq_peak}
P(x,t)=\frac{D_{\text{dis}}}{D_0}x^{-1}e^{-x/\sqrt{D_0 t}}+\frac{D_{\text{dis}}}{D_0}x^{-1}e^{-x^2/4D_ct}\;\mathcal{O}(\sqrt{\frac{D_ct}{x^2}}) . 
\end{equation}
Since $D_c$ is small for large $N$, the first term dominates $P(x,t)$ even for quite small $x$, which appears as a sharp peak around the origin shown in the inset of Fig.\ref{fig_ngdd}. The height of the peak, $P(x=0,t)$, is controlled by $D_c$ as
\begin{eqnarray}
P(x=0,t)&=&\frac{D_\text{dis}}{D_0}\frac{1}{\sqrt{\pi D_ct}}-\frac{D_\text{dis}}{D_0}\frac{1}{\sqrt{D_0 t}}\nonumber\\
& &+\mathcal{O}(\sqrt{D_c/D_0}),
\end{eqnarray}
which is finite for any finite $D_c$ and diverges when $D_c\rightarrow 0$.
Appendix \ref{apdx_px} provides the full derivation on $P(x,t)$. 
%\begin{eqnarray}
%P(x,t)&=&\frac{D_{\text{dis}}}{D_0}x^{-1}\left[e^{-x/\sqrt{D_0 t}}\right.\nonumber\\
%& &\left.+\frac{2}{\sqrt{\pi}}\frac{\sqrt{D_c t}}{x}e^{-x^2/(4D_c t)}+\mathcal{O}(D_c^{3/2})\right].
%\end{eqnarray}

%\begin{eqnarray}
%F&=&\frac{D_{\text{dis}}}{2D_0}x^{-1}\left[e^{-\frac{x}{\sqrt{D_0 t}}}\left(\text{Erf}(\sqrt{D/D_0}-x/(2\sqrt{D t}))\right)\right]
%\end{eqnarray}

\section{Trajectory-based data analysis}
\label{sec_da}

\subsection{Reconstruct the diffusion map from trajectories}
\label{sec_dm}

The structural information of the environment is often represented in the style of diffusion map (D-map), 
which is a map of local diffusivity retrieved from trajectories. D-map is generally constructed in the coarse-grained fashion,
since the trajectories are sparse in most experiments. We say a map is of $s$-resolution, if it is composed of the grains of size $s\le L$. The local diffusivity of each grain is estimated from all the segments centered in the grain $q$ and its nearest-neighbours by
\begin{equation}
\label{eq_ds}
\delta^{(s)}_q=\frac{1}{2d}\frac{1}{K}\sum_{k=1}^{K}\frac{\left\vert x_k(t+dt)-x_k(t)\right\vert^2}{dt},
\end{equation}
where  $dt$ is the time bin and $\{x_k(t)\rightarrow x_k(t+dt)\}$ is the $k$th segment of all the $K$ relevant ones. 
The fidelity of D-map can be evaluated from the mean value of the retrieved local diffusivities averaged over all the $K$ grains in the sample,
\begin{equation}
\bar{\delta}^{(s)}=\frac{1}{M}\sum_{q=1}^{M}\delta^{(s)}_q.
\end{equation}
Fig.\ref{fig_scheme} provides an illustration of the coarse-graining scheme. 

\begin{figure}
\centering
\includegraphics[width=8.6cm]{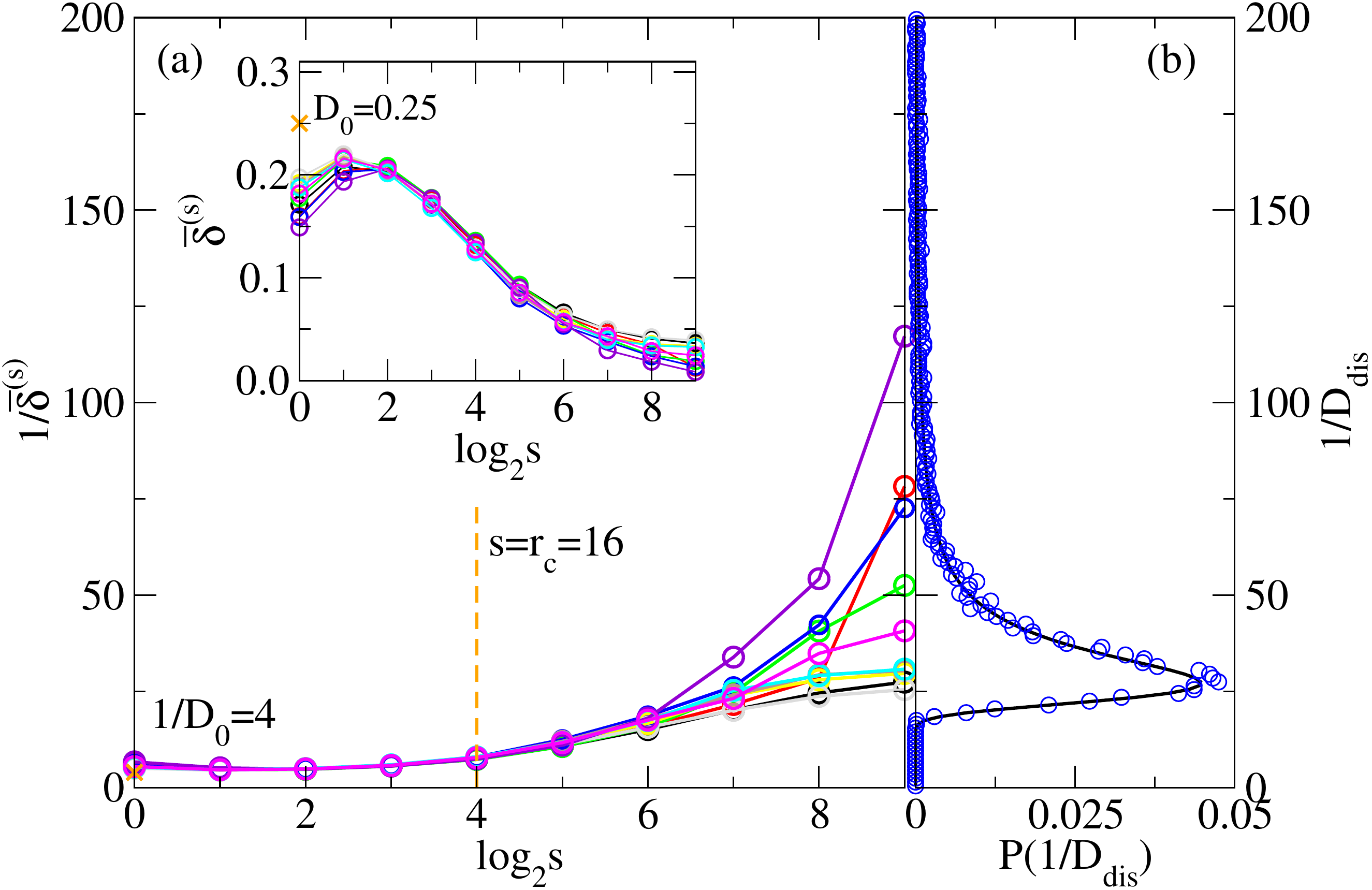}
\caption{\label{fig_drec} (left)The inverse of mean local diffusivity of coarse-grained diffusion map $\left(\bar\delta^{(s)}\right)^{-1}$ for various $s$. The colors denote different disordered samples. The orange cross mark shows $1/D_0$, which is expected for the genuine map. 
(Inset) $\bar\delta^{(s)}$ for various resolutions and different samples. 
(Right) The distribution of $\left(D_{\text{dis}}\right)^{-1}$ obtained by sampling $8000$ disordered realizations (circles). It follows one-sided L\'{e}vy stable distribution with $\mu=1$ (solid line). }
\end{figure}

The inset of Figure \ref{fig_drec} shows $\bar{\delta}^{(s)}$ retrieved for $10$ disordered samples, where $10^4$ trajectories are simulated for each sample. As can be seen in the figure, the fidelity is merely guaranteed for $s<r_c$ that $\bar{\delta}^{(s)}\approx D_0$. $\bar{\delta}^{(s)}$ deviates from $D_0$ in the limit $s\rightarrow 1$, since $10^4$ trajectories are still not enough for reconstruction in the finest resolutions. For coarse-grained maps with $s>r_c$, the deviance becomes more significant, accompanied by the rise of sample-to-sample fluctuation which is clearly presented in the fashion of $\left(\bar{\delta}^{(s)}\right)^{-1}$ (see Fig.\ref{fig_drec}(a)).  Noticing the segments sample the landscape by the Boltzmann weight, one can show $\delta^{(s)}_q$ equals the diffusion coefficient $D_{\text{dis}}^{(s)}$ of the of grain $q$, given by Eq.(\ref{eq_dtau2}) with $N=s^2$ traps. 
In the most coarse-grained case $s=L$, $\delta^{(L)}$ is exactly the diffusion coefficient $D_{\text{dis}}$ estimated from ensemble-averaged MSD of all the trajectories. One can expect that $\left(\bar{\delta}^{(L)}\right)^{-1}$ follows the stable distribution provided by Eq.(\ref{eq_stable}) (shown in Fig.\ref{fig_drec}(b)). $\bar{\delta}^{(s)}$ introduces a path linking the local diffusivity and the instantaneous diffusivity, from determined $\bar{D}^{(l)}=D_0$ to random $\bar{D}^{(t)}=D_{\text{dis}}$, which is shown by color lines in Fig.\ref{fig_drec}(a). 

\subsection{Estimate the correlation length}
\label{sec_cl}
It is worthy to notice the sample-to-sample fluctuation of $\left(\bar{\delta}^{(s)}\right)^{-1}$ is depressed in D-map with fine resolutions. It leads to an approach to estimate the typical correlation length in quenched samples as follows: 
\begin{enumerate}[1)]
\item Divide the whole sample into sub-samples of size $s$.
\item Measure diffusion coefficient $\delta^{(s)}_q$ of each sub-sample $q$ and calculate the mean value $\bar{\delta}^{(s)}\equiv\frac{1}{M}\sum_{q=1}^M \delta^{(s)}_q$ over all the $M$ sub-samples. 
\item Repeat 1) and 2) for various $s$, and record the $s$ dependence of $\bar\tau^{(s)}=\left(\bar\delta^{(s)}\right)^{-1}$. 
\item Repeat 1), 2) and 3) for various samples, calculate the sample-to-sample fluctuation $\Delta^2\equiv\left<(\bar\tau^{(s)})^{2}\right>-\left<\bar\tau^{(s)}\right>^2$.
\end{enumerate}
Fig.\ref{fig_scheme} shows the above procedure. 
In the case the $x^{-1}$-peak is observed in the non-Gaussian distribution of displacement, 
one may also expect that $\Delta^2$ falls for decreasing $s$ as that shown in Fig.\ref{fig_drec}(a). For the diffusion map with $s<r_c$, each sub-sample is dominated by a unique local diffusivity. One may expect $\Delta^{2}$ vanishes. The random samples are hence characterized by two universal parameters, the correlation length $r_c$ and the mean local diffusivity $\bar\delta^{(s)}\vert_{s<r_c}\simeq D_0$. 

\section{Discussion}
\label{sec_ds}

We provide here a brief reasoning for the sharp peak in $P(x,t)$. 
Due to inadequate trajectories, data from particle tracking experiments is usually analyzed in the manner of time averaging. 
In the time ensemble, additional sampling weight joins in the statistics, which appears in this work as the Boltzmann weight (Eq.(\ref{eq_boltzmann})) depending on the local diffusivity (Eq.(\ref{eq_px})). The sharp peak hence arises. Similar phenomena have been reported as ``population splitting'' in aging continuous time random walk (CTRW)\cite{barkai2003prl,barkai2003jcp,schulz13,schulz14}. In this annealed model, sub-diffusion is expected and the displacement distribution $P(x,t)$ is in general non-Gaussian. The aging CTRW is hence a convenient model for ``sub-diffusive and non-Gaussian'' phenomena. We notice that Cherstvy \textit{et al.} have also investigated various types of anomalous diffusions in static environments with deterministic local diffusivity\cite{cherstvy13, cherstvy14}. In this work, we show the sharp peak also exists in the ``Brownian, but non-Gaussian'' diffusion while the environment is disordered and static. 

The main results in Sec.\ref{sec_ue} and Sec.\ref{sec_da} largely depend on the heavy tail of the sojourn time distribution, $P(\tau)\sim\tau^{-2}$, which is contributed by the most deeply trapped particles.
 The $\tau^{-2}$ tail of $P(\tau)$ leads to the L\'evy stable distribution with exponent $\mu=1$ (Eq.(\ref{eq_stable})), which is the marginal case that $\ln N$ correction appears. There have been recently theoretical progresses on the subtle and important case \cite{burov2012, zarfaty2017}. We would also like to note that the diffusion is not exactly Brownian due to the $\ln$-correction, although it might be ignored in experiments. 
 
 In particle tracking experiments, it is a common observation that a portion of particles are pinned at the initial positions over the whole experiment. The threshold $D_c$ for the single particle diffusivity $D_s$ is widely employed, smaller than which the particle is labelled ``immobile''. (See e.g. [9].) A small-$D$ cutoff is hence introduced in the statistics over the ``mobile'' trajectories. It again leads to Eq.(\ref{eq_pdt1}), but from practical consideration in experiments. We notice that the threshold $D_c$ is typically quite small. $D^{-1}\exp[-D/D_0]$ behavior would still dominate $P(D^{(t)})$ observed in experiments, such as that in \cite{he16}.

In this work, the spatially correlated local diffusivity $\{D^{(l)}_i\}$ is generated from the extreme landscape $\{V_i\}$ by $D^{(l)}_i=e^{V_i/T}/2d$.  The $T=1$ case leads to exponential distributed $\{D^{(l)}_i\}$, which was reported by experiments. In more general $T\neq 1$ cases, $P(D^{(l)}_i)$ could be stretched or compressed as $P(D^{(l)}_i=D)= D_0^{-1}TD^{T-1}e^{-D^T/D_0}$. The whole class of models provide description of a range of non-Gaussian diffusion in static disordered environments. 

\section{Summary}
\label{sec_sm}
In summary, we have introduced a quenched trap model for non-Gaussian diffusion in a static disordered environment, 
which is spatially correlated. This model is largely inspired by the particle tracking experiments on non-Gaussian diffusion, especially those of more crowded and static media such as cytoplasm of ATP-depleted cells\cite{munder16} and cell membrane\cite{he16,jeon16}. The relaxation time of such media would be longer than the observation time. The assumption of static disorder may hence apply. Our analytical and numerical studies show a localization mechanism due to the coupling between the local diffusivity and the sojourn time in the trap, which leads to population splitting as a sharp peak in $P(x,t)$ around $x=0$. 
%We note that $P(D^{(t)})$ reported by the previous experiment on cell membrane\cite{he16} is of the same shape predicted by our model. 
Cells would largely benefit from this phenomenon since the biological functions of molecular machines are mostly carried out by the immobile ones. We analytically estimate the diffusion coefficient and its fluctuation among disordered samples. Due to the heavy tail of sojourn time distribution, the diffusion coefficient is depressed by the sample size. The size-dependent effect of $D_{\text{dis}}$ calls our attention to the fidelity of the coarse-grained diffusion map, which is a widely used approach to visualize the structure infromation obtained from particle tracking experiments. Our study suggests the fidelity is guaranteed only in the case that the spatial resolution is high enough to identify the fine structures in the disordered environment. On the other hand, it offers us an approach to estimate the typical correlation length, from the trajectories in a large bunch of samples. We hope this work would shed a light on the experiments on cells where the cell-to-cell fluctuation is always significant. 

\begin{acknowledgements}
We enjoyed inspiring discussions with Hui Li, Fangfu Ye, Penger Tong and Lei-Han Tang. We are very grateful to Erik Aurell and Eli Barkai for comments. This work is supported by National Natural Science Foundation of China (Grant No. 11705064, 11675060,  91730301),
Fundamental Research Funds for the Central Universities (Grant No. 2662015QD005), and the Huazhong Agricultural University Scientific and Technological Self-innovation Foundation Program (Grant No.2015RC021). 
\end{acknowledgements}

\appendix

\section{Negative moments of one-sided L\'{e}vy stable distribution with exponent $\mu=1$}
\label{apdx_nm}
Let $x_N$ follow $S(x;1,1,\gamma,\delta_N)$ with $\delta_N=\delta_0+\frac{2}{\pi}\gamma\ln N$. 
The negative moments of $x_N$ is defined as
\begin{equation}
\left<x_N^{-\alpha}\right>=\int_{0}^{\infty}dx\;x^{-\alpha}S(x;1,1,\gamma,\delta_N),\text{  }\alpha>0.
\end{equation} 
Because the concerned distribution is not strictly stable, the moments depend on $N$.  
In this section, we estimate the large-$N$ dependence of $\left<x_N^{-\alpha}\right>$. 

The approach is similar to that by Chechkin {\it et al.}\cite{chechkin09}, where they deal with the $\mu<1$ case. 
Noting
\begin{equation}
x^{-\alpha}=\frac{1}{\Gamma(\alpha)}\int_0^{\infty}ds\;s^{\alpha-1}e^{-xs}, \alpha>0, 
\end{equation}
we have
\begin{eqnarray}
\label{eq_xa}
\left<x_N^{-\alpha}\right>&=&\frac{1}{\gamma^{\alpha}}\left<\left(\frac{x_N}{\gamma}\right)^{-\alpha}\right>\nonumber\\
	&=&\frac{1}{\gamma^{\alpha}}\int_0^{\infty}dx\;x^{-\alpha}f(x)\nonumber\\
	&=&\frac{1}{\gamma^{\alpha}\Gamma(\alpha)}\int_0^{\infty}ds\;s^{\alpha-1}\int_0^{\infty}dx\;e^{-xs}f(x)\nonumber\\
	&=&\frac{1}{\gamma^{\alpha}\Gamma(\alpha)}\int_0^{\infty}ds\;s^{\alpha-1}\overline{f}(s),
\end{eqnarray}
where $f(x)\equiv P(\frac{x_N}{\gamma}=x)$ and its Laplace transform $\overline{f}(s)=\mathcal{L}(f(x))$.
Noting the Laplace transform of $g(x)\equiv P(\frac{x_N}{\gamma}-\delta_N=x)=S(x;1,1,1,0)$ is given by \cite{uchaikin99}
\begin{equation}
\overline{g}(s)=\mathcal{L}\left(S(x;1,1,1,0)\right)=e^{-s\ln s},
\end{equation}
we can see
\begin{equation}
\overline{f}(s)=\mathcal{L}(f(x))=\mathcal{L}(g(x-\delta_N))=e^{-s(\delta_N+\ln s)}.
\end{equation}

For the large-$N$ dependence of $\left<x_N^{-\alpha}\right>$, we estimate the integral in Eq.(\ref{eq_xa}). 
The integral can be separated into three parts:
\begin{eqnarray}
I&=&I_1+I_2\nonumber\\
&=&\left(\int_0^{1}ds\;+\int_1^{\infty}ds\;\right)Q(s),
\end{eqnarray}
where the integrand $Q(s)=s^{\alpha-1}e^{-s(\delta_N+\ln s)}$.

For the first part, we can see
\begin{equation}
\int_0^{1}ds\;s^{\alpha-1}e^{-s\delta_N}<I_1<\int_0^{1}ds\;s^{\alpha-1}e^{-s\delta_N-\min(s\ln s)}.
\end{equation}
which gives 
\begin{equation}
\frac{\Gamma(\alpha)-\Gamma(\alpha,\delta_N)}{\delta_N^{\alpha}}<I_1<e^{1/e}\frac{\Gamma(\alpha)-\Gamma(\alpha,\delta_N)}{\delta_N^{\alpha}},
\end{equation}
where $\Gamma(\alpha,x)=\int_x^{\infty}dt\;t^{\alpha-1}e^{-t}$ is the incomplete gamma function. 
Since $\Gamma(\alpha,x)\sim x^{1-\alpha}e^{-x}$, it can be neglected for large $N$. 
We see $I_1$ is controlled by $N$ as
\begin{equation}
I_1\sim \delta_N^{-\alpha}.
\end{equation}

For the second part, we have
\begin{equation}
0<I_2<\int_1^{\infty}ds\;s^{\alpha-1}e^{-s\delta_N}=\frac{\exp(-\delta_N)}{\delta_N}(1+O(\delta_N^{-1})).
\end{equation}

For large $N$, $I_1$ dominates the whole integral $I$. 
The negative moments $\left<x_N^{-\alpha}\right>$ depends on $N$ as
\begin{equation}
\left<x_N^{-\alpha}\right>\sim \left(\delta_0+\frac{2}{\pi}\ln N\right)^{-\alpha}.
\end{equation}
It is confirmed by numerical sampling, as shown in Fig.3 of the main text. 

\section{The distribution of instantaneous diffusivity $P(D^{(t)})$}
\label{apdx_dt}

In this appendix, we estimate $P(D^{(t)}=D\vert D_{\text{dis}})$ given by Eq.(\ref{eq_pdt0}) as
\begin{eqnarray}
\label{eq_dtd}
P(D^{(t)}=D\vert D_{\text{dis}})&=&\sum_{i=1}^N \left[P(x=x_i\vert D_i^{(l)}=D, D_{\text{dis}})\times\right.\nonumber\\
& &\left. P(D_i^{(l)}=D\vert D_{\text{dis}})\right].
\end{eqnarray}
The conditional probability of the position is given by Eq.(\ref{eq_px}) as
\begin{equation}
\label{eq_xd}
P(x=x_i\vert D_i^{(l)}=D, D_{\text{dis}})=\frac{D_{\text{dis}}}{N}D^{-1}.
\end{equation}
We focus on the estimation of the conditional probability $P(D_i^{(l)}\vert D_{\text{dis}})$. 
The Bayesian law suggests
\begin{equation}
\label{eq_dld}
P(D_i^{(l)}\vert D_{\text{dis}})=P(D_i^{(l)})\frac{P(D_{\text{dis}}\vert D_i^{(l)}) }{P(D_{\text{dis}})}.
\end{equation}
One may notice that $D_{\text{dis}}$ depends on $D^{(l)}_i$ via $D_{\text{dis}}=\left[\frac{1}{N}\sum_{j=1} (D_j^{(l)})^{-1}\right]^{-1}$.  A decoupled $D_{\text{dis}}'$ can be hence introduced as
\begin{equation}
\label{eq_d1}
D_{\text{dis}}'=\left[\frac{1}{N-1}\sum_{j\neq i} (D_j^{(l)})^{-1}\right]^{-1},
\end{equation}
which is independent of $D_i^{(l)}$. 
Its relation to $D_{\text{dis}}$ is given by
\begin{equation}
D_{\text{dis}}'=\frac{N-1}{N}\frac{1}{1-\epsilon}D_{\text{dis}}.
\end{equation}
where we denote $\epsilon\equiv D_{\text{dis}}/(N D^{(l)}_i)$ for short. 
Noting 
\begin{equation}
P(D_{\text{dis}}\vert D_i^{(l)})dD_{\text{dis} }=P(D'_{\text{dis}}\vert D_i^{(l)}) dD'_{\text{dis}},
\end{equation}
we can see
\begin{equation}
\label{eq_disd}
P(D_{\text{dis}}\vert D_i^{(l)})=\frac{1}{(1-\epsilon)^{2}}P(D'_{\text{dis}}=\frac{D_{\text{dis}}}{1-\epsilon}). 
\end{equation}
Noting Eq.(\ref{eq_d1}), one can see $P(D'_{\text{dis}})=0$ for any $D'_{\text{dis}}<0$. It leads to a natural cutoff for $\epsilon>1$, which gives
\begin{equation}
\label{eq_dld0}
P(D_{\text{dis}}\vert D_i^{(l)})=0, \text{ for }\epsilon=\frac{D_{\text{dis}}}{N D^{(l)}_i}>1. 
\end{equation}
It is to say, $D^{(l)}_i$ is bounded by $D_c=D_{\text{dis}}/N$ from below. 
Using Eq.(\ref{eq_dtd}), Eq.(\ref{eq_dld}), Eq.(\ref{eq_dld0}), we arrive at Eq. (\ref{eq_pdt1}) in the main text
\begin{equation}
\label{eq_apdt1}
P(D^{(t)}=D\vert D_{\text{dis}})=0, \text{ for }D\le D_c. 
\end{equation}
In the region $\epsilon\ll 1$,  Eq.(\ref{eq_disd}) suggests 
\begin{equation}
\label{eq_apdt2}
P(D_{\text{dis}}\vert D_i^{(l)})\approx P(D_{\text{dis}}), \text{ for }D\gg D_c. 
\end{equation}
That is to say $D_{\text{dis}}$ is independent of the shallow traps $D^{(l)}_i\gg D_c$. 
Noting also Eq.(\ref{eq_dtd}-\ref{eq_dld}) and Eq.(\ref{eq_pd}), we arrive at Eq. (\ref{eq_pdt2}) in the main text
\begin{equation}
P(D^{(t)}=D\vert D_{\text{dis}})=\frac{D_{\text{dis}}}{D_0}D^{-1}e^{-D/D_0}, \text{ for }D\gg D_c.
\end{equation}

\section{The distribution of displacement $P(x,t)$ }
\label{apdx_px}
In this appendix, we estimate $P(x,t)$ given by Eq.(\ref{eq_peak0}) as a convolution over $D^{(t)}$
\begin{equation}
P(x,t)=\int_{0}^{\infty}dD\;G(x,t\vert D^{(t)})P(D^{(t)}=D\vert D_{\text{dis}}).
\end{equation}
Noting that $D^{(t)}$ is bounded from below by $D_c$ (Eq.(\ref{eq_apdt1})) and the independent approximation
(Eq.(\ref{eq_apdt2})), the above convolution becomes
\begin{equation}
P(x,t)=\int_{D_c}^{\infty}dD\; \frac{1}{\sqrt{4\pi D t}}e^{-x^2/4 D t}\frac{D_{\text{dis}}}{D_0}D^{-1}e^{-D/D_0}.
\end{equation}
The primitive function of the integral can be expressed as
\begin{eqnarray}
F(D)&=&\frac{D_{\text{dis}}}{2D_0}\frac{1}{x}\left[e^{-x/\sqrt{D_0 t}}\left(\text{erf}(\sqrt{\frac{D}{D_0}}-\frac{x}{2\sqrt{Dt}})-1\right)\right.\nonumber\\
& &\left.+e^{x/\sqrt{D_0 t}}\left(\text{erf}(\sqrt{\frac{D}{D_0}}+\frac{x}{2\sqrt{Dt}})-1\right)\right],
\end{eqnarray}
where $\text{erf}(x)$ is the error function. It is easy to see $F(D=\infty)=0$. 
We hence arrive at $P(x,t)=-F(D=D_c)$.
For $x\gg\sqrt{D_c t}$ and $D_c\ll D_0$, it can be expanded as
\begin{eqnarray}
P(x,t)&=&\frac{D_{\text{dis}}}{D_0}\frac{1}{x}\left[e^{-x/\sqrt{D_0 t}}+\frac{2}{\sqrt{\pi}}\frac{\sqrt{D_c t}}{x}e^{-x^2/4D_c t}\right.\nonumber\\
& &\left.+\mathcal{O}(D_c^{3/2})\right],
\end{eqnarray}
which is dominated by the first term as 
\begin{equation}
P(x,t)\simeq \frac{D_{\text{dis}}}{D_0}x^{-1}e^{-x/\sqrt{D_0 t}}, \text{ for } x\gg\sqrt{D_c t}. 
\end{equation}
Considering the height of the peak at $x=0$, we have
\begin{eqnarray}
P(x=0,t)&=&\int_{D_c}^{\infty}dD\;\frac{1}{\sqrt{4\pi D t}}\frac{D_{\text{dis}}}{D_0}D^{-1}e^{-D/D_0}\nonumber\\
&=&\frac{1}{\sqrt{4\pi D t}} \left[\frac{1}{\sqrt{D_c t}}\frac{1}{\sqrt{\pi}}e^{-D_c/D_0}\right.\nonumber\\
& &\left.-\frac{1}{\sqrt{D_0 t}}\text{erfc}\left(\sqrt{\frac{D_c}{D_0}}\right)\right],
\end{eqnarray}
where $\text{erfc}(x)$ is the complemental error function. 
In experiment practice, it is a usual case that $D_c\ll D_{\text{dis}}<D_0$. The expansion around $D_c/D_0\sim0$ gives
\begin{eqnarray}
P(x=0,t)&=&\frac{D_{\text{dis}}}{D_0}\frac{1}{\sqrt{\pi}}\left[\frac{1}{\sqrt{D_ct}}-\frac{1}{\sqrt{D_0 t}}\sqrt{\pi}\right.\nonumber\\
& &\left.+\frac{1}{\sqrt{D_0 t}}\mathcal{O}\left(\left(\frac{D_c}{D_0}\right)^{1/2}\right)\right]. 
\end{eqnarray}
For $D_c>0$, the height of the peak is finite and controlled by $1/\sqrt{D_c t}$ as
\begin{equation}
P(x=0,t)\simeq \frac{D_{\text{dis}}}{D_0}\frac{1}{\sqrt{\pi}}\frac{1}{\sqrt{D_c t}}, 
\end{equation}
which decays over time as $t^{-1/2}$.


\begin{thebibliography}{99}

\bibitem{barkai12} E.~Barkai, Y.~Garini and R.~Metzler, Phys. Today \textbf{ 65}, 29 (2012).

\bibitem{bressloff13} P.~C.~Bressloff and J.~M.~Newby, Rev.~Mod.~Phys. \textbf{ 85}, 135 (2013).

\bibitem{wang09} S.~C.~Bae, B.~Wang, J.~Guan and S.~Granick, Proc.~Natl.~Acad.~Sci. \textbf{106}, 15160 (2009).

\bibitem{wang12} B.~Wang, J.~Kuo, S.~C.~Bae and S.~Granick, Nat.~Mater. \textbf{11}, 481 (2012).

\bibitem{kou17} B.~Kou, et al., Nature \textbf{551}, 360 (2017).

\bibitem{stylianidou14} S.~Stylianidou, N.~J.~Kuwada and P.~A.~Wiggins, BioPhys. J. \textbf{107}, 2684 (2014).

\bibitem{li15} H.~Li, S.-X. Dou, Y.-R. Liu, W.~Li, P.~Xie, W.-C. Wang and P.-Y. Wang, J. Am. Chem. Soc. \textbf{137}, 436 (2015).

\bibitem{munder16} M.~C.~Munder et al., eLife \textbf{5}, e09347 (2016).

\bibitem{he16} W.~He, H.~Song, Y.~Su, L.~Geng, B.~J.~Ackerson, H.~B.~Peng and P.~Tong, Nat. Commun. \textbf{7}, 11701 (2016).
 
\bibitem{jeon16} J.-H.~Jeon, M.~Javanainen, H.~Martinez-Seara, R.~Metzler and I.~Vattulainen, Phys. Rev. X \textbf{6}, 021006 (2016).
 
\bibitem{zhao16} J.~Wang, Y.~Zhang and H.~Zhao, Phys. Rev. E \textbf{ 93}, 032144 (2016).

\bibitem{ghosh16}  S.~K.~Ghosh, A.~G.~Cherstvy, D.~Grebenkov, R.~Metzler, New J. Phys. \textbf{ 18},  013027 (2016).
 
\bibitem{metzler13} A.~G.~Cherstvy, A.~V.~Chechkin and R.~Metzler, New J. Phys. \textbf{15}, 083039 (2013).
 
\bibitem{slater14} M.~V.~Chubynsky and G.~W.~Slater, Phys. Rev. Lett. \textbf{113}, 098302 (2014).

\bibitem{uneyama15} T.~Uneyama, T.~Miyaguchi and T.~Akimoto, Phys.~Rev.~E \textbf{ 92}, 032140 (2015). 
 
\bibitem{akimoto16pre} T.~Akimoto and E.~Yamamoto, Phys. Rev. E \textbf{93}, 062109 (2016).
 
\bibitem{miyaguchi16} T.~Miyaguchi and T.~Akimoto and E.~Yamamoto, Phys. Rev. E \textbf{94}, 012109 (2016).

\bibitem{bodrova16} A.~S.~Bodrova, A.~V.~Chechkin, A.~G.~Cherstvy, H.~Safdari, I.~M.~Sokolov and R.~ Metzler, Sci. Rep. \textbf{ 6}, 30520 (2016).

\bibitem{cherstvy16} A.~G.~Cherstvy and R.~Metzler, Phys. Chem. Chem. Phys. \textbf{ 18}, 23840 (2016). 

\bibitem{aurell17} E.~Aurell and S.~Bo, Phys. Rev. E \textbf{96}, 032140 (2017).

\bibitem{manzo15} C.~Manzo, J.~A.~Torreno-Pina, P.~Massignan, G.~J.~Lapeyre, Jr. , M.~Lewenstein and M.~F.~Garcia Parajo, Phys. Rev. X \textbf{5}, 011021 (2015).

\bibitem{metzler17} A.~V.~Chechkin, F.~Seno, R.~Metzler and I.~M.~Sokolov, Phys.~Rev.~X \textbf{7}, 021002 (2017).

\bibitem{ghosh15} S.~K.~Ghosh, A.~G.~Cherstvy, R.~Metzler, Phys. Chem. Chem. Phys. \textbf{ 17}, 1847 (2015).

\bibitem{ye12} B.-S.~Lu, F.~Ye, X.~Xing, P.~M.~Goldbart, Phys. Rev. Lett. \textbf{108}, 257803 (2012).

\bibitem{haus87} J.~W.~Haus and K.~W.~Kehr, Phys.~Rep. \textbf{150}, 263 (1987).

\bibitem{bouchaud90} J.~P.~Bouchaud and A.~Georges, Phys.~Rep. \textbf{ 195}, 127 (1990).

\bibitem{monthus03} C.~Monthus, Phys.~Rev.~E \textbf{ 67}, 046109 (2003).

\bibitem{bertin03} E.~M.~Bertin and J.~P.~Bouchaud, Phys.~Rev.~E \textbf{ 67}, 026128 (2003).

\bibitem{barkai11} S.~Burov and E.~Barkai, Phys.~Rev.~Lett. \textbf{106}, 140602 (2011). 

\bibitem{luo14} L.~Luo and L.-H.~Tang, Chin.~Phys.~B \textbf{ 23}, 070514 (2014).

\bibitem{luo15} L.~Luo and L.-H.~Tang, Phys.~Rev.~E \textbf{ 92}, 042137 (2015).

\bibitem{akimoto16} T.~Akimoto, E.~Barkai and K.~Saito, Phys.~Rev.~Lett. \textbf{ 117}, 180602 (2016).

\bibitem{luo16} L.~Luo and M.~Yi, Sci. China-Phys.~Mech.~Astron.~ \textbf{ 59}, 120521 (2016).

\bibitem{nolan} J.~Nolan, {\it Stable Distribution: Models for Heavy-Tailed data}, book in progress, 2014. 

\bibitem{chechkin09} A.~V.~Chechkin, M.~Hofmann and I.~M.~Sokolov, Phys.~Rev.~E \textbf{80}, 031112(2009).

\bibitem{uchaikin99} V.~V.~Uchaikin and V.~M.~Zolotarev, \textit{Chance and stability. Stable distributions and their applications} (Walter de Gruyter 1999) doi:10.1515/9783110935974
 
\bibitem{gillespie77} D.~T.~Gillespie, J.~Phys.~Chem. \textbf{ 81}, 2340 (1977).


\bibitem{barkai2003prl} E.~Barkai, Phys.~Rev.~Lett. \textbf{ 90}, 104101 (2003)

\bibitem{barkai2003jcp} E.~Barkai and Y.-C. Cheng, J.~Chem.~Phys. \textbf{ 118}, 6167 (2003).

\bibitem{schulz13} J.~H.~P.~Schulz, E.~Barkai and R.~Metzler, Phys.~Rev.~Lett. \textbf{ 110}, 020602 (2013).

\bibitem{schulz14} J.~H.~P.~Schulz, E.~Barkai and R.~Metzler, Phys.~Rev.~X \textbf{ 4}, 011028 (2014).

\bibitem{cherstvy13} A.~G.~Cherstvy and R.~Metzler, Phys.~Chem.~Chem.~Phys. \textbf{ 15}, 20220 (2013).

\bibitem{cherstvy14} A.~G.~Cherstvy and R.~Metzler, Phys. Rev. E \textbf{ 90}, 012134 (2014). 

\bibitem{burov2012} S.~Burov and E.~Barkai, Phys.~Rev.~E \textbf{ 86}, 041137 (2012).

\bibitem{zarfaty2017} L.~Zarfaty, A.~Peletskyi, I.~Fouxon, S.~Denisov and E.~Barkai,  	arXiv:1712.04397 (2017). 
 
\end{thebibliography}
\end{document}